# Arm Tangents and the spiral structure of the Milky Way – the Age Gradient

Jacques P Vallée

Herzberg Astronomy and Astrophysics, National Research Council of Canada

ADDRESS   5071 West Saanich Road, Victoria, British Columbia, Canada V9E 2E7
ORCID    http://orcid.org/0000-0002-4833-4160
EMAIL    jacques.p.vallee@gmail.com

**Abstract.**   From the Sun, a look at the edge of each spiral arm in our Milky Way (seen tangentially, along the line of sight) can yield numerous insights. Using different arm tracers (dust, masers, synchrotron emission, CO gas, open star clusters), we observe here for the first time an age gradient (*about 12 ±2 Myrs/kpc*), much as predicted by the density wave theory.  This implies that the arm tracers are leaving the dust lane at a relative speed of *about 81 ±10 km/s.*  We then compare with recent optical data obtained from the Gaia satellite, pertaining to the spiral arms.

## 1.  Introduction

Nowadays, we are at an exciting stage of getting to know better the inside of spiral arms, their origins in space and time, and how they maintain themselves in view of existing localised shocks and turbulences. Since 2006, the trigonometric 'parallax' measurement of radio masers (through interferometric telescopes) is ongoing - the inverse of the parallax gives a *precise* physical distance, provide the error bar in the parallax value is small.  Since about 2014, catalogues of arm tangents, and statistical analyses of these catalogues of observational data, allow us to get the *precise* galactic longitude where an arm tracer (hot or cold dust, hot or cold gas, young or old stars, etc) peaks in intensity as a telescope drifts across the arm in the Galactic plane.  Since 2017, the recent Gaia satellite data have allowed the cartography and velocity of nearby stars in the Galactic disk.   A **Gaia** stellar image of the location of the spiral arms near the Sun showed a diffuse broad area for the Perseus arm and for the Sagittarius arm, with the well-defined narrow arm spine indicated by the radio masers (Fig.1 in Kounkel [1]; Fig.3 in Khopersky [2]; Fig. 8 in Cantat-Gaudin  [3]).

**Figure 1** shows an artist's impression of the disk of the Milky Way.  The huge blue plane disk shows four long curving spiral arms (composed of stars, gas and dust,  cosmic-rays, etc).

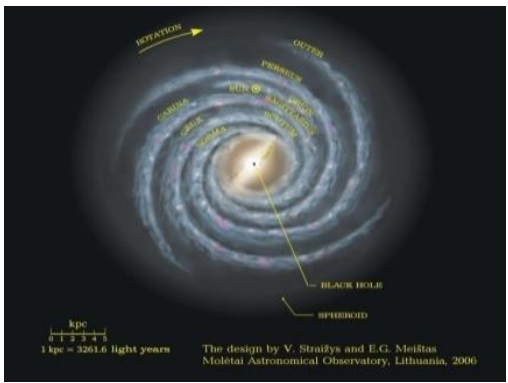

*Figure 1.  The whole view, as would be seen from above the Galactic disk (at the Galactic North Pole). This is* a cartoon drawn by V.Straizys and E.G.Meistas in 2006, at the Lithuania's Moletai Astronomical Observatory, inspired by a drawing published earlier by Vallée  [4].

**Section 2** investigates the 4 long spiral arms, cartographically and kinematically.  **Section 3** presents new quantitative results on the age gradient within a spiral arm, using arm tracers at different ages. **Section 4** concludes.

## 2.  The galactic arms in the Milky Way.

Arms are defined by their location, composition (arm tracers), velocity, pitch, arm start, and width.

**2.1 Cartographic mapping of the Galactic disk.** In the 18th century, scientists with telescopes envision the Center of the Galaxy to be away from the center of the Earth, shattering the predominant centric view (Earth had been thought earlier to be at the center of our Universe). Astronomers like Herschel, Shapley, Oort, IAU commissions used scientific instruments that were improved with time, and finally maser trigonometry to nail the distance to the Galactic Center to a higher precision. Recent observations show that the Sun's orbit is separated from the Galactic Center by about 8.1 kiloparsecs – see Abuter [5].

**Figure 2** shows the coordinate systems accepted nowadays. The direction from the Sun to the Galactic Center (i.e., the 'Galactic Meridian') is set at a Galactic longitude of zero. Galactic longitudes turn counterclockwise around the Sun, until we reach 360 degrees (thus back towards the Galactic Center). A **first** coordinate system is centered on the Sun, to map our disk galaxy: a star is measured with a solar distance r and a galactic longitude l, starting from the vertical axis downward. A **second** coordinate system is centered on the Galactic Center, to map our disk galaxy: a star is measured with a distance to galactic center R and a galactic angle q, starting from the horizontal axis.

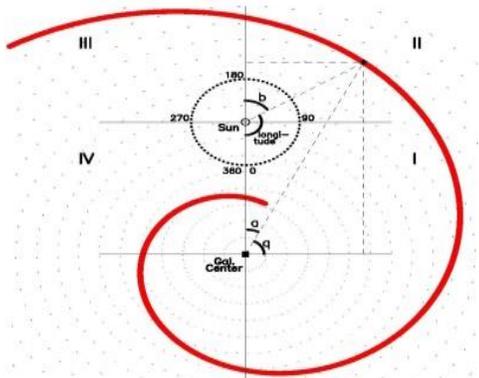

FIGURE 2. Two Map coordinates systems. A black star is shown at top right in a red spiral arm, using two coordinate systems. Galactic quadrants are shown (I: down right; II: top right; III: top left; IV: bottom left). Adapted from Fig. 2 in [6].

Pointing a telescope on Earth to one spiral arm, one can measure the exact galactic longitude when a tracer is 'tangent to the spiral arm' (i.e., when a line of sight from the Sun is aligned along the longest distance inside a spiral arm). One can repeat a scan in galactic longitude to find the exact arm longitude, using each time a different arm tracer (dust, gas, hydrogen, star, etc), thus producing a catalogue of arm tangents. The arms are well measured with geometric mapping, velocity mapping, and future positions are predicted with kinematical models. **Figure 3** shows the observed arm tangent, as seen from the Sun, in CO light, toward the Scutum spiral arm (dashed line). Also shown is the observed arm tangent in 2.4μm dust (dashed line). The offset between the two arm tangents is real, and the dust arm tangent is closer to the direction of the Galactic Center.

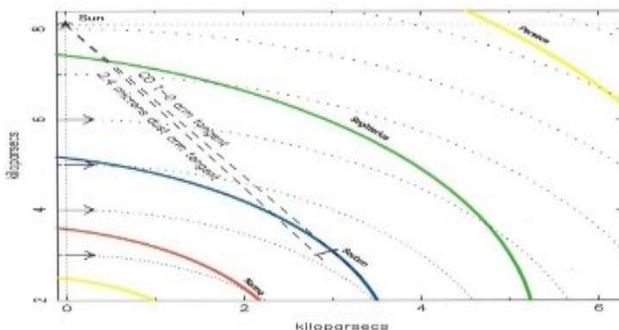

Figure 3. *Spiral arm model, for Galactic Quadrant I, fitted to the cold broad diffuse CO 1-0 gas. The mean longitude of the CO arm tangent from the Sun down to the Scutum arm (upper dashed line) differs by about 2° from the mean longitude for the dust 2.4μm arm tangent (lower dashed line). The short black crosscut at the arm tangents, near (x,y) = (3.0,3.0 kpc), is about 300 pc long. There, the line of sight inside the Scutum arm is shorter for the CO arm tangent (about 1 kpc) than for the dust arm tangent (about 2 kpc). Gas flow (arrows, at left) are going clockwise on a circular orbit around the Galactic Center (dotted circles), entering the arm on its inner arm edge.*

These long spiral arms appear to follow a rough logarithmic shape.  An arm is said to have a 'pitch angle', given by the angle between a tangent to a pure orbital circle (around the Galactic Centre) and the arm's own extended shape.  Looking at the same spiral arm, we can observe the arm tracer as the arm turns around the galactic nucleus, crossing the Galactic Meridian (the straight line from the Sun's location to the center of the Galactic nucleus). First we observe the galactic longitude ($l_I$) in Galactic Quadrant I (where the Sun is traveling to) and second we observe the galactic longitude ($l_{IV}$) in Galactic Quadrant IV (where the Sun is coming from), for the same spiral arm.

**Figure 4** shows a view from above, looking at the galactic disk. Two tangents are drawn from the Sun, one to the Sagittarius arm (bottom right) and one to the Carina arm (bottom left). Carina and Sagittarius are parts of the same arm.   Similarly, two tangents are drawn from the Sun, one to the Scutum arm (to the right) and one to the Crux-Centaurus arm (bottom left), both parts of the same arm.  The difference between galactic longitudes  of two tangents from the same arm yields the pitch angle (p), via trigonometry (see derivation in [6] – equations 1 to 10):

$$(l_I - l_{IV} + \pi) \cdot \tan(p) = \ln(\sin(l_I) / \sin(2\pi - l_{IV}))$$

where π denotes a half-circle, and the galactic longitudes $l_I$ and $l_{IV}$ must be tangents to the same arm tracer (both in CO, say).

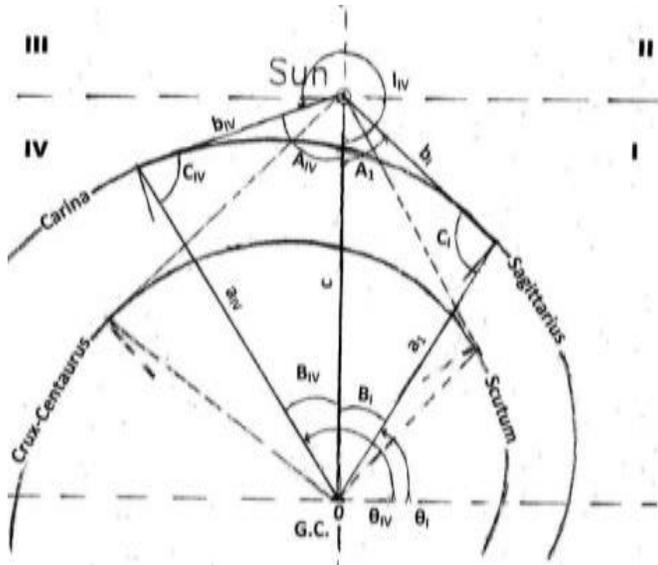

FIGURE 4 –   The arm pitch (deviation from a tangent to a pure circle)   as measured across two galactic quadrants (I and IV). Adapted from Figure 1 in [6].

From Galactic Quadrant I to Galactic Quadrant IV, a logarithmic function was fitted to the Sagittarius-Carina arm, to the Scutum-Crux-Centaurus arm [6], and to the Norma arm [7], all yielding an arm pitch near -13 degrees (downward from a circle).

The four long spiral arms are observed to start near the Galactic Center, but not exactly there: it starts at about 2.2 kpc away from the Galactic Center [8]. Thus the **Perseus** arm is seen to originate near galactic longitude 337° (= -23°) for the cold diffuse broad CO peak, and 340° degrees (= -20°) for the hotter dust lane.  The origin of the **Sagittarius** arm is near galactic longitude 343° (= -17°) for the cold diffuse broad CO peak, and 346° (= -14°) for the masers/young stars lane. The origin of the **Norma** arm is near galactic longitude 016° for the masers/young stars lane,  and 020°    for the cold diffuse broad CO peak.  The **Scutum** arm is seen  to originate near galactic longitude 026° for the dust lane,  and   near 033° for the cold diffuse broad CO peak.

If we measure the arm width 'S'   from the location of the dust lane (not from the maser region) to the location of the diffuse CO gas, then we find a mean value of *315 ± 26 parsecs*  [9]. Here the value of S (signal) over its error (sigma) is 12, thus way above the '5 sigma' threshold normally used in physics to signal a discovery.

**2.2  Velocity/kinematical  mapping the galactic disk**

Centuries ago, the Sun was thought not to move at all. Later commissions of the International Astronomical Union (IAU) results fixed it near 220 km/s. More recently, after several corrections for the Sun's non-circular motion, a final value near 233 km/s was arrived at (Drimmel & Poggio [10]). At the Sun's location, this represents an 'angular velocity' for the material (gas and stars) of about 28.8 km/s/kpc (=233 / 8.1). This value is different than the spiral pattern speed, nearer 17 km/s/kpc [11]. Observations show that the gas and stars move at the same speed, in their roughly circular orbit, around the Galactic Center. They all move clockwise, as seen from the North Galactic Pole.

In our Galaxy, we observe a 'flat' circular orbital curve (i.e., the same circular speed, at every galactic radius), beyond a galactic radius of 2 kpc. Thus, using the known distances to each of the four spiral arms in the Milky Way, a plot of radial velocity versus galactic longitude, or various galactic distances, can be made.

In **Figure 5**, we show the picture towards the interior of the disk, including the Galactic Center. It tells us what a radar on Earth would see when looking towards Galactic quadrant I (at right, longitudes from 0 to 90 degrees) and Galactic quadrant IV (at left, longitudes from -90º or 270º, up to 0º or 360º).

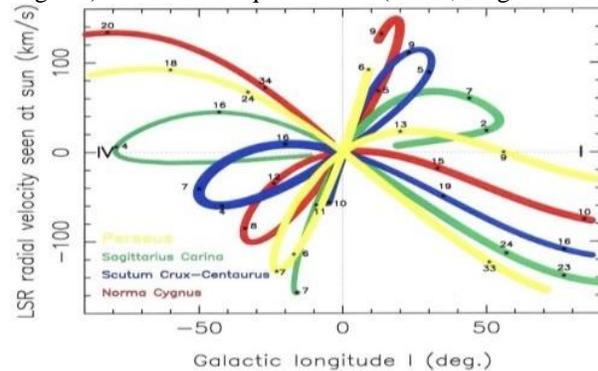

FIGURE 5– The velocity distribution of different spiral arms, looking towards the Galactic center. Adapted from Fig. 3 in [9].

Very similar velocity results were obtained recently by [12] using Gaia EDR3 data in the radial range 5 to 20 kpc, finding a circular orbital velocity near 234.7 ±1.7 km/s near the Sun's position.

The velocity data obtained from the Gaia satellite is making a great contribution to the 3-dimensional spatial structure, and the 3-dimensional velocity distribution of our Galaxy [13]. The passages of the nearby spheroidal Sagittarius Dwarf Galaxy 'SDG' are impacting those stars in the Milky Way disk, to a certain degree, but not enough to explain numerically the plot of $V_z$ versus z [14]; [15]. Gaia DR2 data and other data were used to compute a strong reflex motion of stars in the Galactic disk, due to the last closest approach (50 kpc) of the Large Magellanic Cloud (Fig. 1 in [16]). The encounter between 'SDG' and the Milky Way was confirmed around 540 M years ago [17]. Sysoliatina et al [18] were able to match their star model to the Gaia star data in the solar neighborhood, provided there were two bursts of star formation within the last 4 Gyrs. A recent model [19] argued that the 'phase spiral' ($V_z$ versus z) was excited some 1 to 2 Gyrs before the recent transit of SDG.

The recent **Gaia** satellite has made a contribution to the nearby kinematic structures of our Galaxy ([20]; [21]). They argued that the galactic disk is still in evolution, including a tidal interaction with the incoming Sagittarius dwarf galaxy, a mechanism of compression when stars and gas enter a spiral arm and of expansion when stars and gas exit a spiral arm, and an effect of the rotation of the Galactic bar. One fascinating discovery by Gaia is the apparent 'corrugation'' of the galactic disk, defined loosely as vertical motion ($v_z$) away from the disk (in the z-coordinate); in fact, Antoja et al ([22] – their Fig.1) showed the vertical position and velocities of the stars with Gaia in a plot of $V_z$ versus z, exhibiting a spiral loop. Some explanations of this spiral loop behaviour were made, but the main culprit seems to be due to the passages of the nearby spheroidal Sagittarius Dwarf Galaxy 'SDG' that turns around the Milky Way disk, crossing it at different point in time – see models by Laporte et al [23], Bland-Hawthorn et al [24] and [19]. SDG is at about 16 kpc from our Galactic Center, towards a galactic longitude of 5º and galactic latitude of -15º, with a radial velocity of +140 km/sec, and a size of about 3 kpc ([25] – their Fig. 3); it has been noted for leaving a debris trail in the Local Arm ([26] – their Section 6.4).

Open star clusters have been measured by Gaia; the peak age in the distribution is 3.6 Gyrs, while the shortest age is 3.2 Myrs, and the olderst age is 10 000 Myrs (Fig. 2 in [27]). Such a long age make them difficult to stay with the same spiral arm, and to inform on the spiral arm where they formed.

Recent Gaia maps of the locations of O-B stars (upper main sequence) (Figure 1 in Poggio et al 2021 [28]), open star clusters < 100Myrs (Fig.2 in [28]), and young Cepheids <100Myrs (Fig.3 in [28]) show a large spiral arm width (1 to 2 kpc), much larger than expected from radio tracers, and artifacts caused by foreground extinction, and radial orientations (fingers of God oriented to Earth), mixing stars in interarm islands with stars in spiral arms (0.4-

kpc averaging cells). Gaia does not go mapping beyond the Galactic Center, into was is known as the 'Zona Galactica Incognita' [29].

## 3. New results.

Here we quantify the observed age gradients, and look for sustainable theories.

### 3.1 Age gradient seen in spiral arms.

With time, a young star becomes an old star, and its motion keeps going along their galactic circle around the galactic center. The dust location in a spiral arm is often referred to as the '**dust lane**', starting the edge of a spiral arm, on the arm side closest to the Galactic Center. The nursery contains several stars in formation, or 'protostars'. The nursery can be detected, however, at submillimeter radio wavelengths or extreme far infrared wavelengths. Each protostar may harbor 'radio masers' (a part of the process leading from protostar to star).

When travelling inside the arm, after the dust lane and the maser region inside the dust, a traveller may encounter lanes of other arm tracers, such as the location of the radio synchrotron peak intensity, and the location of the radio recombination lines coming from the ionized gas around ionised hydrogen HII regions (HII regions are made of ionized hydrogen gas, surrounding a hot young star coming fresh out of its nursery). When one reaches the outer side of a spiral arm, one enters a region of cold diffuse gas, notably the location of a lot of diffuse Carbon monoxide (CO) gaseous molecules.

For each spiral arm, we thus can take a 'cross-section', going from one arm edge (dust, say) to the other arm edge (diffuse CO gas or old stars). Knowing the linear distance from the Sun to each arm segment seen tangentially, we can convert from angular separation to linear separation.

**Figure 6** shows the picture of a typical cross section of a spiral arm. Shock and star-formation can occur in the red zone (at right), becoming a young protostar (orange zone), then a young star (green zone), and finally an old star (blue zone). Hence an age gradient is seen (called a color gradient, in optical astronomy).

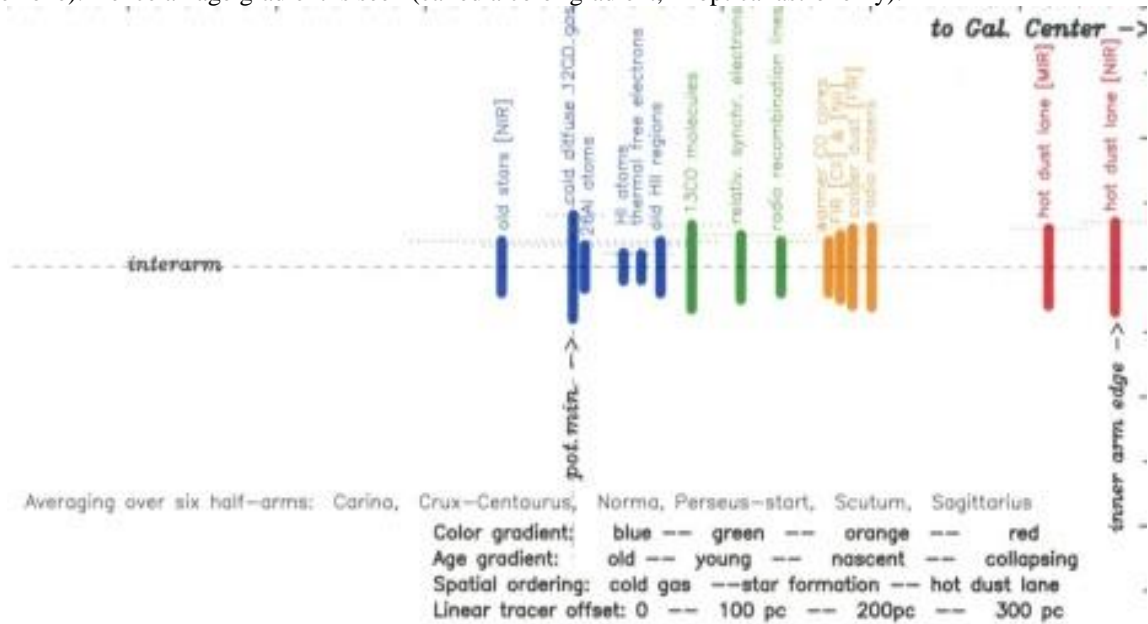

FIGURE 6 – Each arm tracer is located separately in a spiral arm. Statistics over several arms show the location of each arm tracer, in a typical spiral arm cross-section. The distance between the 'potential minimum' (blue zone) and the 'inner arm edge' (red zone) is about 315 pc. Adapted from Fig. 11 in [30].

**Age gradient from different arm tracers.** Given that each tracer peaks in emission at a different age, the observed tracer separation from the dust lane shows an age gradient in the arm. Thus for radio masers and ultracompact HII regions (in the *orange* zone) of typical age 1 Myrs and at about 100 pc from the dust lane and density wave shock (in red zone), the age gradient is about 10±3 Myrs/kpc (=1.0/0.10). For the optically visible young HII regions (in the *green* zone) of typical age 2.2 Myrs and at about 200 pc from the dust lane (in the red zone), the age gradient is about 11±3 Myrs/kpc (=2.2/0.2). For the optically visible old HII regions (in the *blue* zone) of typical age 4 Myrs and at about 300 pc from the dust lane (in the red zone), the age gradient is about 13 ±3

Myrs/kpc (=4/0.3). Elsewhere, the youngest visible optical star clusters (in the *white* area to the left of the blue zone) of typical age 8 Myrs and at about 600 pc from the dust lane would give an age gradient of about 13 ±3 Myrs/kpc (=8/0.6).

We summarise these new results here, by taking a mean over one variable (tracer age over tracer separation from dust lane), giving an age gradient at around 12 ± 2 Myrs/kpc, being around 10 for the orange zone, around 11 for the green zone, around 13 for the blue zone, and around 13 for the white zone. By implication, the relative speed of arm tracers from the shock front (being the inverse of the age gradient) is about 81 ± 10 km/s.

### 3.2 Theories on sustained spiral structure.

Stars and gas go around their own roughly circular orbit around the Galactic Center. Observations showed that the speed on their orbit is the same for everyone, independent of the galactic radius, from a galactic radial distance of 2 to perhaps 20 kpc. In our Galaxy, this fixed speed is near 233 km/sec. Thus the time to do a full orbital circle around the Galactic Center, for the Sun at a galactic radius of 8.1 kpc, is about: $2 \pi$ 8100 pc / 233 km/sec, giving 218 Millions of years.

Many different theories have been published to predict spiral arms (origin, formation, maintenance). Some theories propose that temporary spiral arms can be generated by the 'rotation of a galactic bar' located at the galactic center (predicting 2 spiral arms), or by the 'passage of a nearby galaxy' (predicting 2 tidal spiral arms). Some propose that 'dynamic transient recurrent turbulences' in the galactic disk would create long-lived spiral arms but no age gradient – see a review in Dobbs & Baba (2014) [31] and Pettitt et al (2020) [32]. Others proposed that a long-lived 'spiral density wave', rotating at a fixed pattern speed, could generate 2 or 4 spiral arms (as reviewed in Shu [33] and Roberts [34]).

Density wave theory proposes that a density wave rotates at a fixed angular pattern speed, throughout the galactic disk. Thus at a given galactic radius, the fixed speed of the orbiting gas and stars would differ from that of the pattern speed (given by the galactic radius times the pattern's fixed angular speed). There is a corotation radius (material speed of gas and stars equals the linear speed of the wave pattern). Below that corotation radius, the stars and gas can reach the slower moving wave, creating a 'shock' at the entrance to the arm, which in turn causes the gravitational collapse of pockets of gas into protostars, with their associated masers, the agglomeration of dust into a dust lane, etc. While these protostars are evolving into young stars, they would continue in their orbit, leaving the shock and dust lane behind. As young stars, some would harbor around them an ionised hydrogen region (HII). While these young stars are evolving into old stars, they would continue in their orbit, leaving the young star lane behind.

When proposed by Lin & Shu [35], it was observationally difficult to observe and measure some of their predictions, notably the physical distances with time between protostars, young stars, and older stars, leaving the field open to the creation of alternate theories ([33]; [36]). The predicted separation among different arm tracers inside a spiral arm was first obtained in 2014 for our Milky Way, using statistical analyses of earlier telescopes scans along the galactic plane, in many different arm tracers [37].

At what angular speed does the 'spiral pattern' rotate? Various variants of the density wave theory put it between 15 and 20 km/s/kpc (Table 2 in [38], Vallée [11]). Multiplying by the Sun's distance to the Galactic Center, one finds its linear velocity to be about *145 km/sec*. This value is smaller than the orbital velocity of the gas and stars near the Sun (*233 km/s*), and it ensures that a shock is created as the interarm gas reaches and overtakes a spiral arm.

For the Milky Way (CO gas peak versus dust lane), and for a time T value given by separation over relative speed : T= 315 pc / (233 – 145) km/s = 3.6 Myrs. Thus to see an age gradient, one must use tracers younger than 10 million years. Otherwise, some tracers would have plenty of time to wander away from their initial arm, and reach anywhere over the next spiral arm (confusing their locations there with those born in that next arm). To reach the next spiral arm, away at 0.25x $2 \pi$ 8.1 kpc = 12.7 kpc, and at a relative speed of (233 -145) = 88 km/s, a star needs 144 Myrs.

Summarising here, only the density wave theory seems to fit most observed arm tracers (number of arms, shock and dust, masers and age gradient, gas speed and pattern speed, etc). The others (rotating central bar, passing tides) predict a wrong number of arms or no age gradient (dynamic transient).

## 4.0 Conclusions.

Using arm tangents, an 'age gradient' was found in the Milky Way since 2014, through the combination of several telescopes and the use of statistical analyses of catalogued observational data, giving the ***precise*** galactic longitude of each individual arm tracer (gas, dust, etc) when it peaks in intensity as a telescope drifts across galactic longitudes in the Galactic plane. Finding here the separation of some arm tracers from the dust lane (Section 3.1) yielded an age gradient near $12 \pm 2$ Myrs/kpc, implying that the arm tracers leave the dust lane at a relative speed near $81 \pm 10$ km/s. The density wave theory seems to accommodate most observational data (Section 3.2). Gaia velocity data are helping paint a dynamical picture of the Milky Way. Parallax measurements (radio masers, Gaia optical stars) are helping define the locations of the spiral arms.

**Acknowledgements.** The figure production made use of the PGPLOT software at the NRC Canada in Victoria. I thank Dr James Di Francesco (NRCC in Victoria, BC, Canada) for an in-depth reading of a recent version of this manuscript, and for comments to improve it. I thank V. Straizys and E.G. Meistas for their cartoon of the Milky Way made in 2006.